\documentclass[conference]{IEEEtran}
\usepackage{cite,amssymb}
\usepackage[pdftex]{graphicx}
\usepackage{algorithm}
\usepackage[cmex10]{amsmath}
\usepackage{algorithm}
\usepackage{algorithm}
\usepackage{algpseudocode}
\usepackage{array,balance,epstopdf}

\ifCLASSOPTIONcompsoc
\usepackage[caption=false,font=normalsize,labelfont=sf,textfont=sf]{subfig}
\else
\usepackage[caption=false,font=footnotesize]{subfig}
\fi
\usepackage{url}
\usepackage{amsfonts}
\usepackage{exscale}
\usepackage{relsize}
\usepackage{multicol,color,bm}
\newcommand{\mH}{\mathop{\rm H}}
\newcommand{\mT}{\mathop{\rm T}}
\begin{document}		
\title{A Fast Method for Array Response Adjustment with Phase-Only Constraint}
\author{\IEEEauthorblockN{Xuejing Zhang\IEEEauthorrefmark{1},
		Zishu He\IEEEauthorrefmark{1},
		Bin Liao\IEEEauthorrefmark{2},
		Xuepan Zhang\IEEEauthorrefmark{3}, 
		Yue Yang\IEEEauthorrefmark{1}}
	\IEEEauthorblockA{\IEEEauthorrefmark{1}University of Electronic Science and Technology of China, Chengdu 611731, China}
	\IEEEauthorblockA{\IEEEauthorrefmark{2}Shenzhen University, Shenzhen 518060, China}	
	\IEEEauthorblockA{\IEEEauthorrefmark{3}Qian Xuesen Laboratory of Space Technology, Beijing 100094, China}
	Email: xjzhang7@163.com, zshe@uestc.edu.cn,
	binliao@szu.edu.cn, zhangxuepan@qxslab.cn,
 yueyang@std.uestc.edu.cn}
\maketitle

\begin{abstract}
In this paper, we propose a fast method for array response adjustment with phase-only constraint.
This method can precisely and rapidly adjust the
array response of a given point by only
varying the entry phases of a pre-assigned weight vector.
We show that phase-only
array response adjustment can be formulated as a polygon construction
problem, which can be solved by edge rotation in complex
plain. Unlike the existing approaches,
the proposed algorithm provides an analytical solution
and guarantees a precise phase-only adjustment without pattern distortion.
Moreover, the proposed method is suitable for an arbitrarily given weight vector and has a low computational complexity.
Representative examples are presented to demonstrate the
effectiveness of the
proposed algorithm.
\end{abstract}
\IEEEpeerreviewmaketitle
\section{Introduction}
Controlling the array power response as desired plays an important role in the applications of sensor arrays.
Many methods have been reported to control array response
by designing an architecture including both amplitudes and phases of
excitation for the given specification, see e.g.,
\cite{reff101,2010-5,1992-1,Chan2010}. In fact, phase-only architectures which use a single
power-divider network simplify the beamforming network as well as reduce cost. In this perspective,  phase-only control is preferred  \cite{reff503}. So far, many phase-only
response control and/or pattern synthesis methods have been reported.  A phase-only power synthesis technique
is presented for reconfigurable conformal arrays in \cite{reff016}, whereas for linear and planar arrays,  the phase-only synthesis of minimum peak sidelobe
patterns is considered in \cite{reff702}.
Taking advantage of advances in convex optimization \cite{reff018}, a class of phase-only response control algorithms have been devised.
For instance, the semidefinite relaxation (SDR) technique \cite{reff020}  has been applied in \cite{reff019} and the alternating optimization algorithm is adopted in \cite{reff321}. Other methods include intersection approach \cite{reff011},
bi-quadratic programming method \cite{reff601}, to name just a few.

It is worth noting that in general the aforementioned methods have to redesign the weight vector even if only a slight adjustment
of the array power response is required. To this end, some accurate array response control approaches have been developed recently \cite{reff103,reff102,word}. However, these methods design both the amplitude and phase of the weight vector. They cannot be straightforwardly applied to the phase-only design.
Thus, a geometrical formulation to the problem of phase-only array response adjustment is developed in this paper.
The proposed method is able to realize rapid array response
adjustment with phase-only constraint for a given weight vector.

\section{The Proposed Algorithm}
\subsection{Geometrical Interpretation of the Problem}
We consider an array of $N \geq 3$ elements. For a given weight vector $ {\bf w}_{\rm pre} $ and a pre-assigned angle $ \theta_c $,
we aim to only vary the entry phases of $ {\bf w}_{\rm pre} $ such that the so-obtained new weight vector  $ {\bf w}_{\rm new} $  adjusts the normalized array power response at $ \theta_c $
to its desired level $ \rho_c $, i.e., 
\begin{align}\label{phaseonly001}
L_{\rm new}(\theta_c,\theta_0)\triangleq
{|{\bf w}^{\mH}_{\rm new}{\bf a}(\theta_c)|^2}/{|{\bf w}^{\mH}_{\rm new}{\bf a}(\theta_0)|^2}={\rho_c}
\end{align}
where $\theta_0 $ represents the main beam axis, $ {\bf a}(\theta) $ stands for the steering vector at $ \theta $, and the entries of $ {\bf w}_{\rm new} $ and $ {\bf w}_{\rm pre} $ fulfill the condition
\begin{align}\label{4}
{w}_{{\rm new},n}=|{w}_{{\rm pre},n}|\cdot e^{j\phi_n},~n=1,\cdots,N.
\end{align}
with $ \phi_n=\angle{w}_{{\rm new},n} $. By introducing a phase parameter $ \psi_c \in [0,2\pi) $, we can rewrite \eqref{phaseonly001} as
\begin{align}\label{phaseonly004}
{\bf w}^{\mH}_{\rm new}
\underbrace{(
	{\bf a}(\theta_c)\!-\!\sqrt{\rho_c}e^{j\psi_c}{\bf a}(\theta_0)
	)}_{\triangleq{\bf h}({\theta_c},{\theta_0},\rho_c,\psi_c)}\!=\!\!\sum_{n=1}^{N}h_n|{w}_{{\rm pre},n}|e^{-j\phi_n}\!=\!0
\end{align}
where $ h_n $ is the $ n $th element of $ {\bf h}({\theta_c},{\theta_0},\rho_c,\psi_c) $.
Given $ {\theta_c} $, ${\theta_0}$, $\rho_c$ and $ \psi_c $,
our concern is finding appropriate $ \phi_n $, $n=1,\cdots,N$,
to satisfy \eqref{phaseonly004}.

%

%
%

For notational simplicity, let us define
\begin{align}\label{key2358}
v_n\triangleq
h_n\cdot|{w}_{{\rm pre},n}|,~n=1,\cdots,N
\end{align}
and hence re-express \eqref{phaseonly004} as $\sum_{n=1}^{N}v_ne^{-j\phi_n}=0.$
In a view of complex plane, $ v_ne^{-j\phi_n} $
corresponds to a vector, denoted as
$ \overrightarrow{v_ne^{-j\phi_n}} $,
whose coordinate is given by
$ \left({\Re}(v_ne^{-j\phi_n}),
{\Im}(v_ne^{-j\phi_n})
\right) $.
With this geometrical concept, one can denote \eqref{phaseonly004} as
\begin{align}\label{key035}
\sum_{n=1}^{N}\overrightarrow{v_ne^{-j\phi_n}}=\sum_{n=1}^{N}\overrightarrow{|v_n|e^{j(\vartheta_n-\phi_n)}}=\overrightarrow{\bf 0}
\end{align}
where $ \vartheta_n=\angle{v_n}=\angle{h_n} $, $ n=1,\cdots,N $.
The problem of solving \eqref{phaseonly004} with respect to $\phi_n$ becomes
how to rotate the vectors $ \overrightarrow{v_n} $, $n=1,\cdots, N$,
in complex plane to
sum them up to a zero vector.



\subsection{A Geometric Solution via Triangle Construction}
 We now derive a solution to the above problem via
triangle construction \cite{geo}. To begin with, let us sort the entries of $[|v_1|,\cdots,|v_N|]$ in a  descending order as 
\begin{align}\label{key50}
\left[d_1,\cdots,d_N\right]^{\mT}={\bf J}
\left[|v_1|,\cdots,|v_N|\right]^{\mT}
\end{align}
where  $d_1\geq d_2 \geq \cdots \geq d_N>0$ and  $ {\bf J} $ denotes a certain permutation matrix. As a result, solving the problem \eqref{key035} with respect to $\phi_n$  equals to finding $ \{\varphi_1,\cdots,\varphi_N\} $ such that
\begin{align}\label{key036}
\sum_{i=1}^{N}\overrightarrow{d_ie^{j\varphi_i}}=\overrightarrow{\bf 0}.
\end{align}
After solving this problem with respect to $\varphi_n$, $ n=1,\cdots,N $, the phase $ \phi_n $ can be recovered as
\begin{align}\label{key51}
\left[\phi_1,\cdots,\phi_N\right]^{\mT}\!=
\left[\vartheta_1,\cdots,\vartheta_N\right]^{\mT}\!-{\bf J}^{-1}\!
\left[\varphi_1,\cdots,\varphi_N\right]^{\mT}.
\end{align}

Before proceeding, a useful lemma \cite{geo} is given.
\newtheorem{theoreml}{Lemma} 
\begin{theoreml}
	Assume $d_1\geq d_2 \geq \cdots \geq d_N>0$, define a piecewise summation function $ Q(\cdot) $ as
	\begin{align}\label{key26}
	Q(k,l)\triangleq\sum_{i=k}^{l}d_i,~~~1\leq k\leq l\leq N.
	\end{align}
	Then if
	$ 	d_1\leq Q(2,N) $,
	we have
$ 	d_1\geq\min\limits_{i\in\{2,\cdots,N-1\}}
	|Q(2,i)-Q(i+1,N)| $.
\end{theoreml}

\begin{figure}[!t]
	\centering
	\includegraphics[width=2.9in]{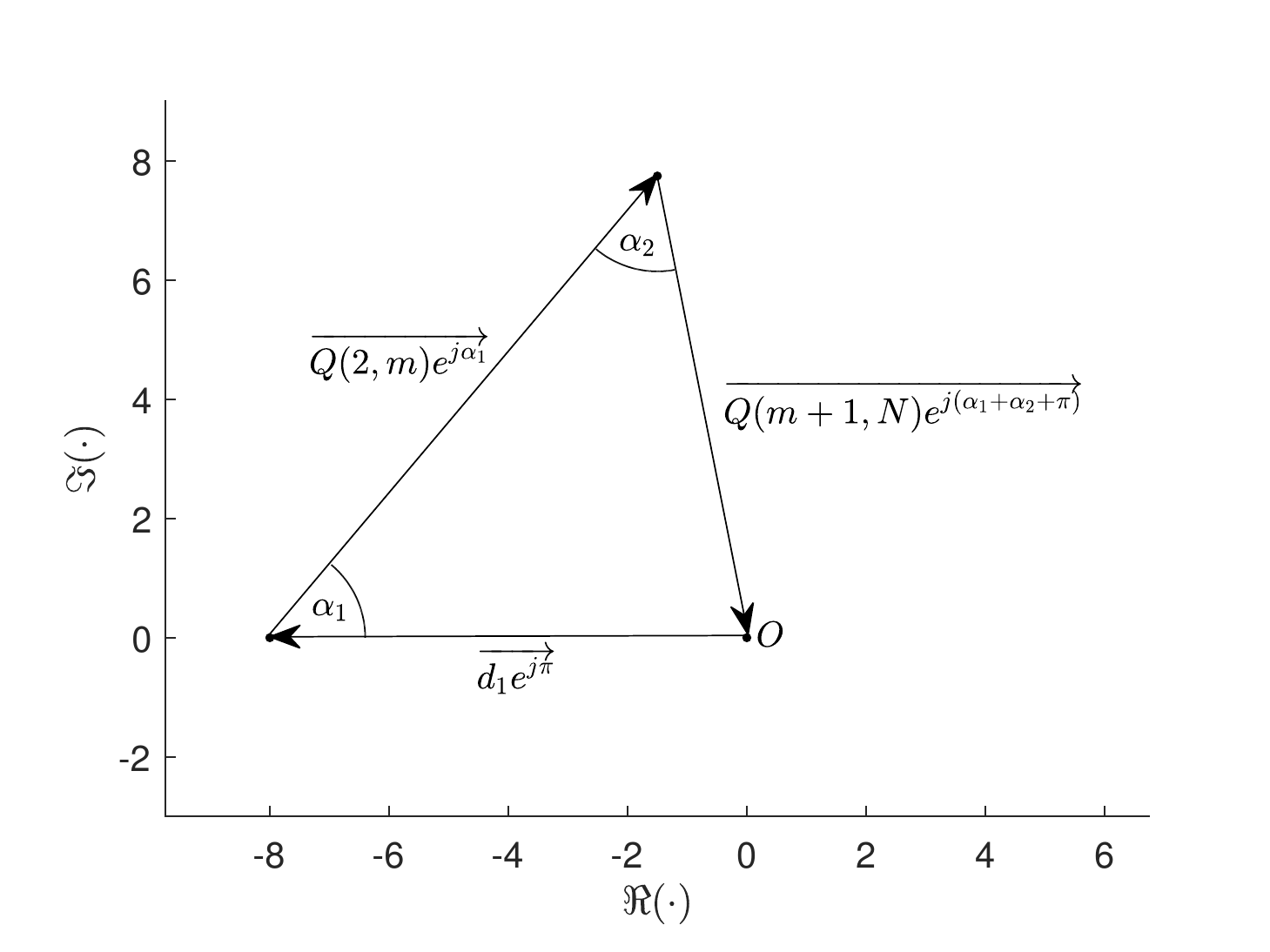}
	\caption{Geometrical illustration of triangle construction.}
	\label{phaseonlyfig1ok}
\end{figure}

On the basis of Lemma 1, the following important
corollary can be obtained.
\newtheorem{theorem2}{Corollary}
\begin{theorem2}
	If $d_1\leq Q(2,N)$, the non-linear Eqn. \eqref{key036} has the following solution as
	\begin{align}\label{okf073}
	\varphi_i=
	\left\{
	\begin{array}{cc}
	\pi,~&{\rm if}~i=1~~~~~~~~~~~~\\
	{\alpha_1},~&{\rm if}~2\leq i\leq m~~~~~~\\
	{\alpha_1}+{\alpha_2}+\pi,~&{\rm if}~m+1\leq i\leq N
	\end{array}
	\right.
	\end{align}
	where $ m $ is the index satisfying
	\begin{align}\label{key06}
	m={\rm arg}\!\!\min_{i\in\{2,\cdots,N-1\}}
	|Q(2,i)-Q(i+1,N)|
	\end{align}
	$ \alpha_1 $ and $ \alpha_2 $ are given by
	\begin{subequations}\label{alpha12}
		\begin{align}
		\alpha_1&={\rm acos}
		\left(
		\dfrac{d^2_1+Q^2(2,m)-Q^2(m+1,N)}{2d_1Q(2,m)}
		\right)\\
		\alpha_2&={\rm acos}
		\left(
		\dfrac{Q^2(2,m)+Q^2(m+1,N)-d^2_1}{2Q(2,m)Q(m+1,N)}
		\right).
		\end{align}
	\end{subequations}
\end{theorem2}
\begin{IEEEproof}
	Given the $ m $ in \eqref{key06},
	if $ d_1\leq Q(2,N) $,
	we can obtain from Lemma 1 that
$ 	d_1\geq |Q(2,m)-Q(m+1,N)| $.
	Thus, the three edges, i.e., $ d_1 $, $ Q(2,m) $
	and $ Q(m+1,N) $, can construct a triangle as shown in Fig. \ref{phaseonlyfig1ok}.
	In a geometrical manner, we have
	$ \overrightarrow{d_1e^{j\pi}}+
	\overrightarrow{Q(2,m)e^{j\alpha_1}}+
	\overrightarrow{Q(m+1,N)e^{j(\alpha_1+\alpha_2+\pi)}}
	=\overrightarrow{\bf 0} $,
	which indicates that there exists a solution to the problem \eqref{key036} 
	as specified in \eqref{okf073}.
	This completes the proof.
\end{IEEEproof}

Note that although Corollary 1 provides a closed-form solution to he problem \eqref{key036}
with the aid of triangle construction, it is experimentally found that the resulting beampattern of $ {\bf w}_{\rm new} $
may lead to large pattern variations in the uncontrolled region, compared to the previous beampattern.

\subsection{Solution Analysis via Polygon Construction}
In this subsection, the above-obtained solution is analyzed with polygon construction.
Before further discussion, we first give the following lemma, which has also
been reported and proofed in \cite{geo}.

\begin{theoreml}
	Given  $d_1\geq d_2 \geq \cdots \geq d_N>0$, there exists a solution to $\sum_{i=1}^{N}\overrightarrow{d_ie^{j\varphi_i}}=\overrightarrow{\bf 0}$ (or all the edges $ d_i $'s can form a polygon),
	if and only if the condition $d_1\leq Q(2,N)$ is satisfied.
\end{theoreml}


%
%

According to the derivation of Corollary 1,
we can construct a triangle using $ d_1 $, $ Q(2,m) $ and
$ Q(m+1,N) $, as presented in Fig. \ref{phaseonlyfig1ok},
with $ m $ being given in \eqref{key06}.
In fact, a polygon can be remained
if we rotate the vector
$ \overrightarrow{d_2e^{j\varphi_2}} $ with certain angles.
To make this clear,
let us introduce an auxiliary vector $ \overrightarrow{x_2e^{j\gamma_2}} $
(with modulus $ x_2 $ and phase $ \gamma_2 $)
pointing from $ \overrightarrow{\bf 0} $ to
$ \overrightarrow{d_1e^{j\pi}}+\overrightarrow{d_2e^{j\varphi_2}} $.
We know that all the edges $ d_i $'s ($ i=1,\cdots,N $) can form
a polygon if and only if:
\begin{itemize}
	\item[1)] The edges $ d_1 $, $ d_2 $ and $ x_2 $ can form a triangle.
	\item[2)] The edges $ x_2 $, $ d_3 $, $\cdots$, $d_N $ can form a polygon.
\end{itemize}
Recalling Lemma 2, we can then calculate
the range of feasible $ x_2 $ and
further determine
the set of feasible $ \varphi_2 $.
Once $ \varphi_2 $ has been determined as $ \varphi_{2,\star} $ (discussed in the next subsection), the resulting
$ \overrightarrow{x_2e^{j\gamma_{2,\star}}} $ ($ \gamma_{2,\star} $ represents the ultimate selection of $ \gamma_2 $)
satisfies
$ \overrightarrow{x_2e^{j\gamma_{2,\star}}}=\overrightarrow{d_1e^{j{\varphi}_{1,\star}}}+\overrightarrow{d_2e^{j\varphi_{2,\star}}}
$.

In a general sense, if $ \overrightarrow{x_{i-1}e^{j\gamma_{i-1}}} $ has been determined
as $ \overrightarrow{x_{i-1}e^{j\gamma_{i-1,\star}}} $,
we can then calculate the set of feasible $ \varphi_{i} $, by introducing an auxiliary vector $ \overrightarrow{x_ie^{j\gamma_{i}}} $ pointing from $ \overrightarrow{\bf 0} $ to
$ \overrightarrow{x_{i-1}e^{j\gamma_{i-1,\star}}}+\overrightarrow{d_ie^{j\varphi_i}} $, $ i=2,\cdots,N-2 $, where $ x_1 $
and $ \gamma_{1,\star} $ are defined, respectively, as
$ x_1\triangleq d_1 $,
$ \gamma_{1,\star}\triangleq\varphi_{1,\star}=\pi $.
To satisfy \eqref{key036}, the edges $ x_{i-1} $, $ d_i,\cdots,d_N $
should form a polygon.
According to Lemma 2, we can obtain the set of $ x_i $, denoted as $ \mathbb{X}_i = [x_{i,\min}, x_{i,\max}] $, $2\leq i \leq N-2$, with 
\begin{align}
x_{i,\min} &= {\max}\left\{\big|x_{i-1}-d_i\big|, d_{i+1}-\sum\nolimits_{k=i+2}^Nd_k\right\}\\
x_{i,\max} &=  {\min}\left\{ x_{i-1}+d_i,\sum\nolimits_{k={i+1}}^Nd_k 		\right\}.
\end{align}

\begin{figure}[!t]
	\centering
	\includegraphics[width=3.0in]{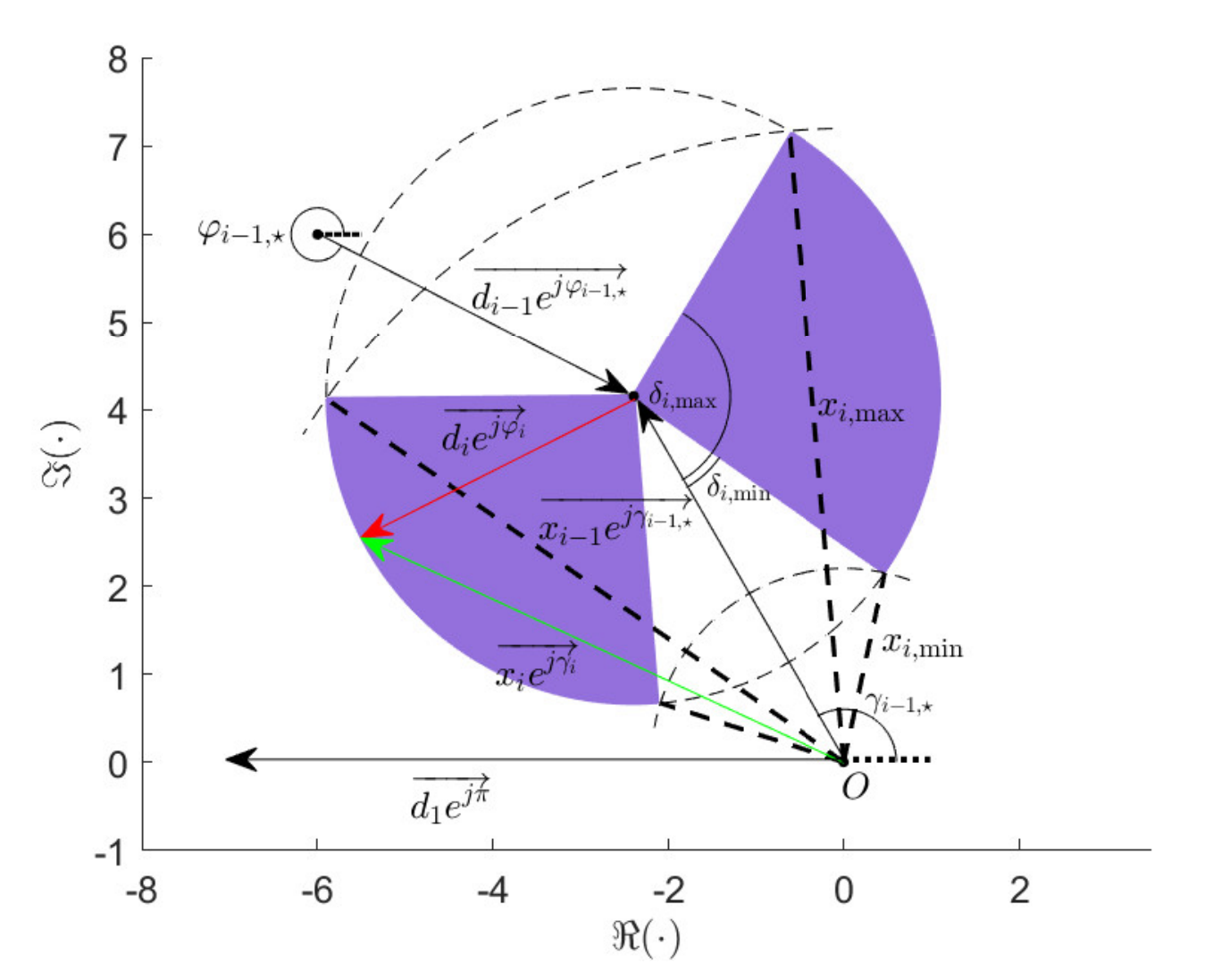}
	\caption{Illustration on geometrical determination of the set of $ \varphi_i $, $ i=2,\cdots,N-2 $.}
	\label{dqiw333}
\end{figure}

Fig. \ref{dqiw333} shows a geometrical interpretation of determining the set of $ \varphi_i $, $2\leq i \leq N-2$.
It is seen that the included angle
between edges $ x_{i-1} $ and $ d_i $ (denoted as $ \delta_i $) satisfies
$ \delta_{i}\in[\delta_{i,{\rm min}},~\delta_{i,{\rm max}}] $
with
\begin{align}
\delta_{i,\rm min}&={\rm acos} \left( ({x_{i-1}^2+d^2_i-x^2_{i,\rm min}})/({2x_{i-1}d_i})\right)\\
\delta_{i,\rm max}&={\rm acos} \left( ({x_{i-1}^2+d^2_i-x^2_{i,\rm max}})/({2x_{i-1}d_i})
\right).
\end{align}
Furthermore, we can obtain the set of feasible $ \varphi_i $ as
\begin{align}\label{key062}
{\varphi_i}\in \Psi_i \triangleq
&[\gamma_{i-1}+\pi-\delta_{i,\rm max},\gamma_{i-1}+\pi-\delta_{i,\rm min}]\cup \nonumber\\
&~~~~[\gamma_{i-1}+\pi+\delta_{i,\rm min}, \gamma_{i-1}+\pi+\delta_{i,\rm max}].
\end{align}
Assume that $ \varphi_i $ is determined as $ \varphi_{i,\star} $ (see next subsection), the ultimate
$ \overrightarrow{x_ie^{j\gamma_{i,\star}}} $, $i=2,\cdots,N-2$, can be expressed
as
\begin{align}
{\overrightarrow{x_{i}e^{j\gamma_{i,\star}}}}={\overrightarrow{x_{i-1}e^{j\gamma_{i-1,\star}}}}+
{\overrightarrow{{d}_{i}e^{j\varphi_{i,\star}}}}.
\end{align}

Note that if $ i=N-2 $ is applied, the resulting
$ x_{N-2} $ can form a triangle with the other two edges
$ d_{N-1} $ and $ d_N $.
In this case, it's not hard to
learn that there are two candidates at most for
$ \varphi_{N-1} $ as
\begin{align}
\varphi_{N-1}\in
\Psi_{N-1}\triangleq\{\gamma_{N-2,\star}+\pi-\delta_{N-1},
\gamma_{N-2,\star}+\pi+\delta_{N-1}\}.\nonumber
\end{align}
where
$ \delta_{N-1}={\rm acos}
\left(
({x_{N-2}^2+d^2_{N-1}-d^2_N})/({2x_{N-2}d_{N-1}})
\right) $.
Moreover,
if $ \varphi_{N-1} $ is
selected as $ \varphi_{N-1,\star} $ (discussed in
the next subsection),
there would be only one choice for the ultimate
$ \varphi_N $
(denoted as $ \varphi_{N,\star} $), which can be
expressed accordingly as
\begin{align}\label{keyvarphiN}
\varphi_{N,\star}=\angle{
	\left(
	{\overrightarrow{x_{N-2}e^{j\gamma_{N-2,\star}}}}+
	{\overrightarrow{{d}_{N-1}e^{j\varphi_{N-1,\star}}}}
	\right)}.
\end{align}

\subsection{Determination of the Phase Parameter $\varphi_i$}
In this subsection, we consider the
determination of $ \varphi_{i,\star} $, $ i=2,\cdots,N $, and complete
the proposed algorithm by finding the ultimate $ \phi_n $, $ n=1,\cdots,N $.

To begin with, for any given $ i\in\{1,\cdots,N\} $,
we note
from \eqref{key51} that
there exists a unique index $ n\in\{1,\cdots,N\} $ such that
$ {\bf J}(i,n)=1 $, and we denote its
resulting value for clarity as
$ n=\varsigma(i) $.
With the above notation, one obtains from \eqref{key51} that
\begin{align}\label{key901}
\phi_{\varsigma(i)}=\vartheta_{\varsigma(i)}-\varphi_i,~i=1,\cdots,N.
\end{align}
Since the beampattern is invariant to a fixed phase shift, we set
the ultimate selection of $ \phi_{\varsigma(1)} $
straightforwardly as
$
\phi_{\varsigma(1),\star}=\vartheta_{\varsigma(1)}-\varphi_{1,\star}=
\vartheta_{\varsigma(1)}-\pi$.

For the index $ i\in\{2,\cdots,N-1\} $,
we propose to
optimize $ \phi_{\varsigma(i)} $ ($ i=2,\cdots,N-1 $) as
\begin{subequations}\label{ob082}
	\begin{align}
	\mathop {\rm minimize}_{\phi_{\varsigma(i)}}
	&~~~|{\rm exp}(j\phi_{\varsigma(i)})-{\rm exp}(j\angle{\overline{w}_{\varsigma(i)}})|\\
	{\rm subject~to}&~~~{\phi_{\varsigma(i)}\in\Phi_{\varsigma(i)}}
	\end{align}
\end{subequations}
where
$\Phi_{\varsigma(i)}=
\left\{\phi_{\varsigma(i)}\big|\phi_{\varsigma(i)}=\vartheta_{\varsigma(i)}-\varphi_i,~
{\varphi_i}\in\Psi_i\right\}$.
In \eqref{ob082}, $ \overline{w}_{\varsigma(i)} $ is
the $ {\varsigma(i)} $th element of a pre-designed weight vector
$ \overline{\bf w} $
that can result a satisfactory beampattern.
Note that the phase-only
constraint \eqref{4} may not be satisfied
by $ \overline{\bf w} $.
In this paper, we  construct
$ \overline{\bf w} $ using the
weight vector orthogonal decomposition (WORD) algorithm \cite{word}.
Giving a weight vector $ {\bf w}_{\rm pre} $, WORD can precisely and analytically adjust array response level at one pre-assigned angle
$ \theta_c $ as the desired level $ \rho_c $. When $ i=N $ is applied, there is only one candidate for $ \varphi_N $, and hence, one can calculate the corresponding
$ \phi_{\varsigma(N),\star} $ as
$ \phi_{\varsigma(N),\star}=\vartheta_{\varsigma(N)}-\varphi_{N,\star} $. Consequently,  the new weight vector $ {\bf w}_{\rm new} $ can be obtained as
\begin{align}
\!\!{\bf w}_{\rm new}\!=\!
\left[
|{w}_{{\rm pre},1}|,\cdots\!,|{w}_{{\rm pre},N}|
\right]^{\mT}\odot
\left[
e^{j\phi_{1,\star}},\cdots\!,e^{j\phi_{N,\star}}
\right]^{\!\mT}\!
\end{align}
where $\odot$ denotes elementwise product. 

\vspace{1em}
\section{Numerical Results}
In this section, simulations are presented to demonstrate the
effectiveness of the proposed algorithm.
For comparison purpose, the results of SDR method in \cite{reff019}
and convex relaxation (CR) method in \cite{reff321} will also be
presented.

\begin{table}[!t]
	\renewcommand{\arraystretch}{1.15}
	\caption{Settings of Sensor Locations, Excitation Amplitudes
		and the Obtained Weightings by
		the Proposed Algorithm}
	\label{table4}
	\centering
	\begin{tabular}{c | c | c | c}
		\hline
		$~ n ~$&$ ~x_n(\lambda)~ $&$ ~~g_n~~ $ & $ ~~~~~~~~w_n~~~~~~~~ $\\
		\hline
		1&0.00 &1.12  	& $ ~~1.1169 + i0.0838 $\\
		2&0.30 &1.10  	& $ ~~0.5680 - i0.9420 $\\
		3&0.90 &1.00 	& $ -0.9167 - i0.3996 $\\
		4&1.55 &1.05  	& $ ~~0.0797 + i1.0470 $\\
		5&2.05 &0.98 	& $ ~~0.9661 - i0.1647 $\\
		6&2.60 &1.06   	& $ -0.4119 - i0.9767 $\\
		7 &3.05 &0.91 	& $ -0.8384 - i0.3538 $\\	
		8 &3.60 &0.95 	& $ ~~0.3063 + i0.8993 $\\
		9 &4.05 &1.02 	& $ ~~1.0014 - i0.1938 $\\
		10&4.55 &0.92  	& $ -0.0239 - i0.9197 $\\
		11&5.00 &0.98  	& $ -0.9758 + i0.0910 $\\		
		\hline
	\end{tabular}
\end{table}

\begin{figure}[!t]
	\centering
	\includegraphics[width=3.3in]{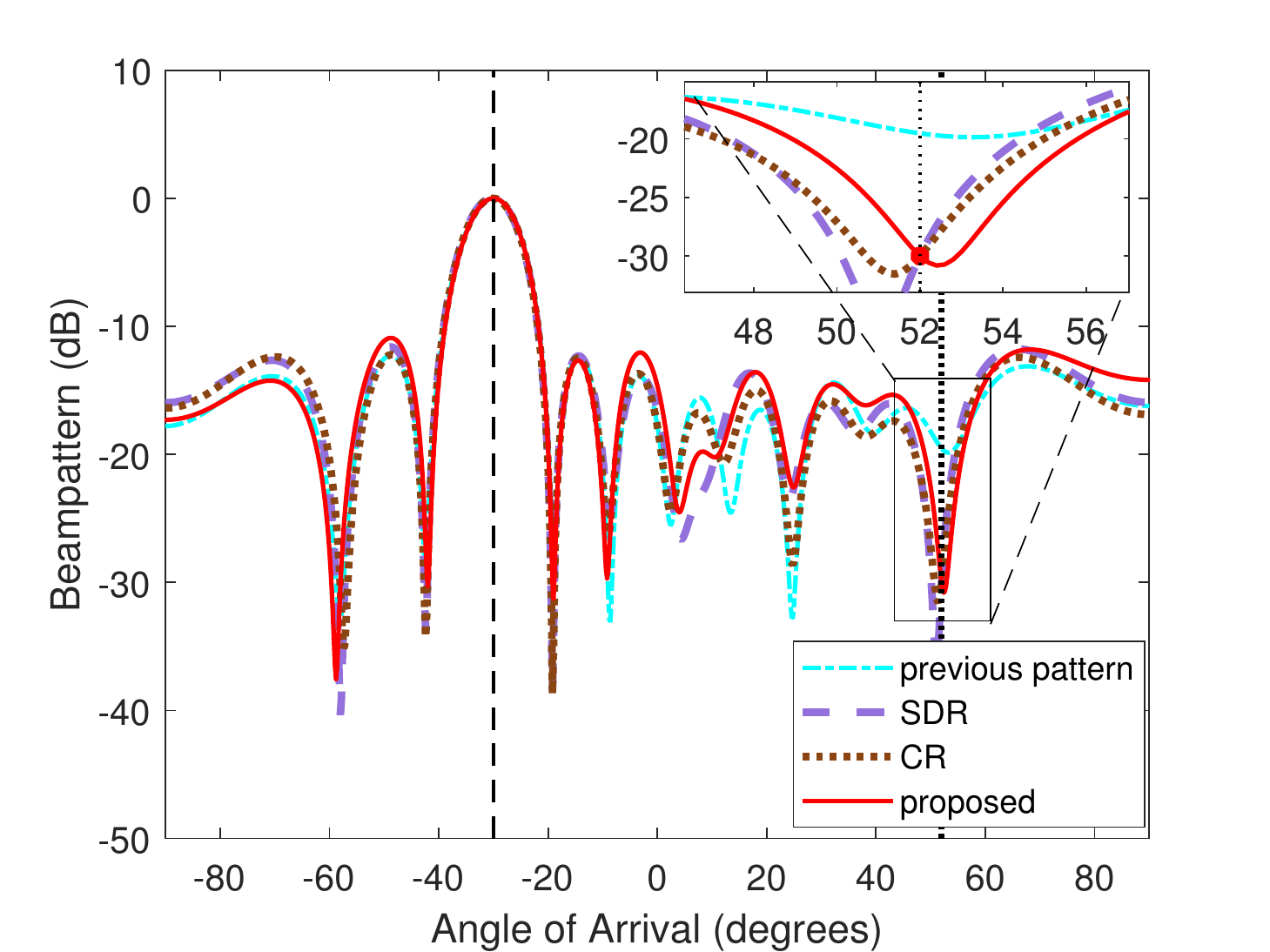}
	\caption{Resulting patterns using a non-uniform linear array.}
	\label{NonULAFromQui}
\end{figure}

\subsubsection{Phase-Only Array Response Adjustment for a Non-uniformly Spaced Linear Array}
In the first example, we consider a non-uniformly spaced linear array
with 11 isotropic elements, see $ x_n $ in Table \ref{table4} for the
specifications of sensor positions.
The main beam axis is set as $ \theta_0=-30^{\circ} $.
The controlled angle and its desired level are taken as $ \theta_c=52^{\circ} $
and $ \rho_c=-30{\rm dB} $, respectively.
In this case, the previous weight $ {\bf w}_{\rm pre} $
is prescribed as
$ {\bf w}_{\rm pre}={\bf g}\odot{\bf a}(\theta_0) $,
see $ g_n $ in Table \ref{table4} for its $ n $th entry specification.
Fig. \ref{NonULAFromQui} presents the resulting beampatterns
of different methods with the above configuration.
We can see that all the three methods adjust
the response of $ \theta_c $ to its desired level $ \rho_c $.
However, the CR method doesn't realize a complete phase-only control.
The resulting weight of the proposed algorithm is listed in Table \ref{table4}, and the corresponding beampattern brings small
pattern variations at the uncontrolled region as shown in Fig. \ref{NonULAFromQui}.

\begin{figure}[!t]
	\centering
	\includegraphics[width=3.3in]{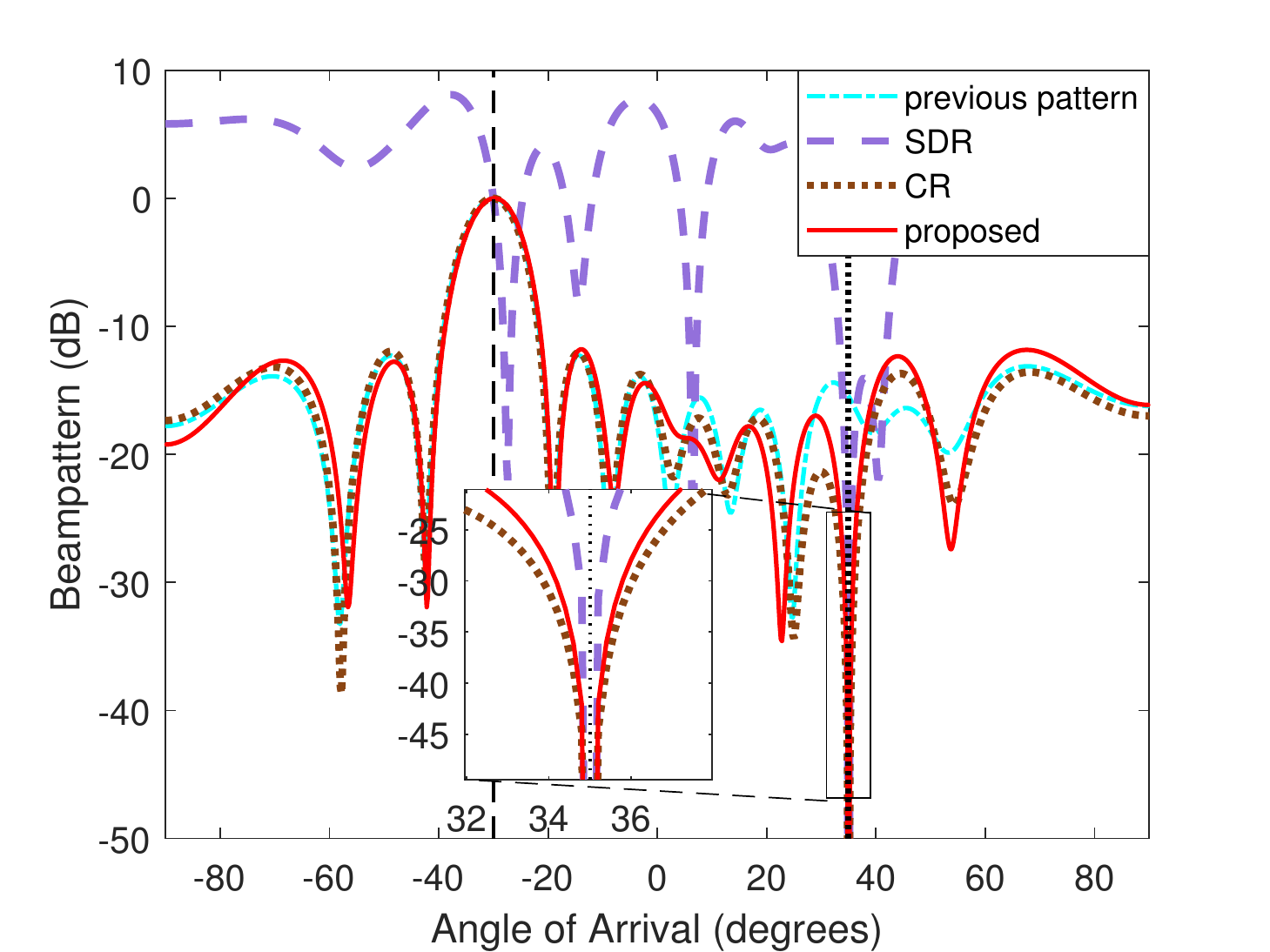}
	\caption{Resulting patterns with phase-only nulling.}
	\label{NULL}
\end{figure}

\subsubsection{Phase-Only Nulling}
Following the array configuration and the setting of $ {\bf w}_{\rm pre} $
in the preceding example,
we keep $ \theta_0=-30^{\circ} $ and attempt to form a null at
$ \theta_c=35^{\circ} $.
The resulting beampatterns of different approaches are depicted in Fig. \ref{NULL}.
One can clearly see that all the three methods shape a deep notch
at $ \theta_c $.
Moreover, the CR method and the proposed one result
small pattern variations at the uncontrolled region.
On the contrary, the ultimate pattern of SDR method is distorted seriously.
The possible reason for this adverse consequence is
the semidefinite relaxation operator, which reformulates a
convex problem that is not equivalent to the original one.
In this case, the proposed algorithm can realize a phase-only control, which may
not be always guaranteed by the CR method.
In addition, the execution time of
the proposed algorithm is
great shorter than those of SDR and CR.

\section{Conclusion}
In this paper, we have presented a
geometrical approach to fast array response adjustment
with phase-only constraint.
The devised method can precisely and rapidly adjust the response of a given point
with the phase-only constraint, staring from a pre-assigned weight vector.
In the proposed algorithm, the single-point phase-only
array response adjustment is recast as a polygon construction
problem, which can be solved by edge rotation.
Representative simulations have been carried out to verify
the effectiveness and superiority of the proposed algorithm.

\bibliography{radarConference20181110}
\end{document}